\documentclass[aps,prl,showpacs,twocolumn,superscriptaddress]{revtex4}

\usepackage[dvips]{graphicx}
\usepackage{epsfig}
\usepackage{dcolumn}
\usepackage{bm}
\usepackage{latexsym}
\usepackage{amssymb}
\usepackage{amsfonts}
\usepackage{amsmath}
\usepackage{color}

\usepackage[colorlinks,bookmarks=false,citecolor=blue,linkcolor=red,urlcolor=blue]{hyperref}
\usepackage{verbatim}

\def\be{\begin{equation}}
\def\ee{\end{equation}}

\def\vec{\mathbf}
\def\mc{\mathcal}

\begin{document}

\title{Spin-Peierls instabilities of antiferromagnetic rings in a magnetic field}
\author{Valeria Lante}\email{valeria.lante@uninsubria.it}
\affiliation{Dipartimento di Fisica e Matematica, Universit\`a dell'Insubria, Via Valleggio 11, I-22100 Como, Italy}
\affiliation{Institut de th\'eorie des ph\'enom\`enes physiques, Ecole Polytechnique F\'ed\'erale de Lausanne, CH-1015 Lausanne, Switzerland}
\author{Ioannis Rousochatzakis}
\affiliation{Institut de th\'eorie des ph\'enom\`enes physiques, Ecole Polytechnique F\'ed\'erale de Lausanne, CH-1015 Lausanne, Switzerland}
\author{Karlo Penc} 
\affiliation{Research Institute for Solid State Physics and Optics, H-1525 Budapest, P.O. Box 49, Hungary}
\author{Oliver Waldmann}
\affiliation{Physikalisches Institut, Universit\"at Freiburg, Hermann-Herder-Stra{\ss}e 3, 79104 Freiburg, Germany}
\author{Fr\'ed\'eric Mila}
\affiliation{Institut de th\'eorie des ph\'enom\`enes physiques, Ecole Polytechnique F\'ed\'erale de Lausanne, CH-1015 Lausanne, Switzerland}

\date{\today}
         
\begin{abstract}
Motivated by the intriguing properties of magnetic molecular wheels at field induced level crossings,
we investigate the spin-Peierls instability of  antiferromagnetic rings in a field by exact diagonalizations
of a microscopic spin model coupled to the lattice via a distortion dependent Dzyaloshinskii-Moriya
interaction. We show that, beyond the unconditional instability at level crossings for infinitesimal 
magnetoelastic coupling, the model is characterized by a stronger tendency to distort at higher level crossings,
and by a dramatic angular dependence with very sharp torque anomalies when the field is almost
in the plane of the ring. These predictions are shown to compare remarkably well with available 
torque and Nuclear Magnetic Resonance data on CsFe$_8$. 
\end{abstract}


\maketitle
Intermediate between single spins and bulk magnets, molecular magnets have
attracted a lot of attention since they offer a suitable platform for probing 
the predictions of quantum mechanics, e.g. for the tunneling probability between 
almost ``classical'' states\cite{GSV}. 
At the field-induced level crossings (LCs), a small gap is in most cases 
opened by small anisotropies, allowing the system upon sweeping a magnetic field
to remain in the same state or to tunnel following the adiabatic 
ground state\cite{GSV,Chudnovsky,Chiorescu,V6_PRL}.
The subclass of ring-like molecular magnets is exceptional in that respect. 
The structure of these ``magnetic wheels'' is in general so symmetric that, although 
present, anisotropic interactions are predicted to leave intact the degeneracy at LCs. 
However, this prediction is in contradiction with several experiments.   
The direct evidence of level repulsion in Fe6:Li clusters lead Affronte \textit{et al.}\cite{Affronte} to
postulate a distortion at low temperatures that would allow extra terms such
as Dzyaloshinskii-Moriya (DM) interactions\cite{DM}. Cinti \textit{et al.}\cite{Cinti} have 
introduced a model with rigid dimerization and fixed DM anisotropy to account for the 
tunnel splittings in Fe6:Na clusters,  while Miyashita \textit{et al.}\cite{Miyashita} have 
discussed the possibility of DM anisotropy induced by thermal fluctuations.

\begin{figure}[!b]
\includegraphics[width=0.475\textwidth]{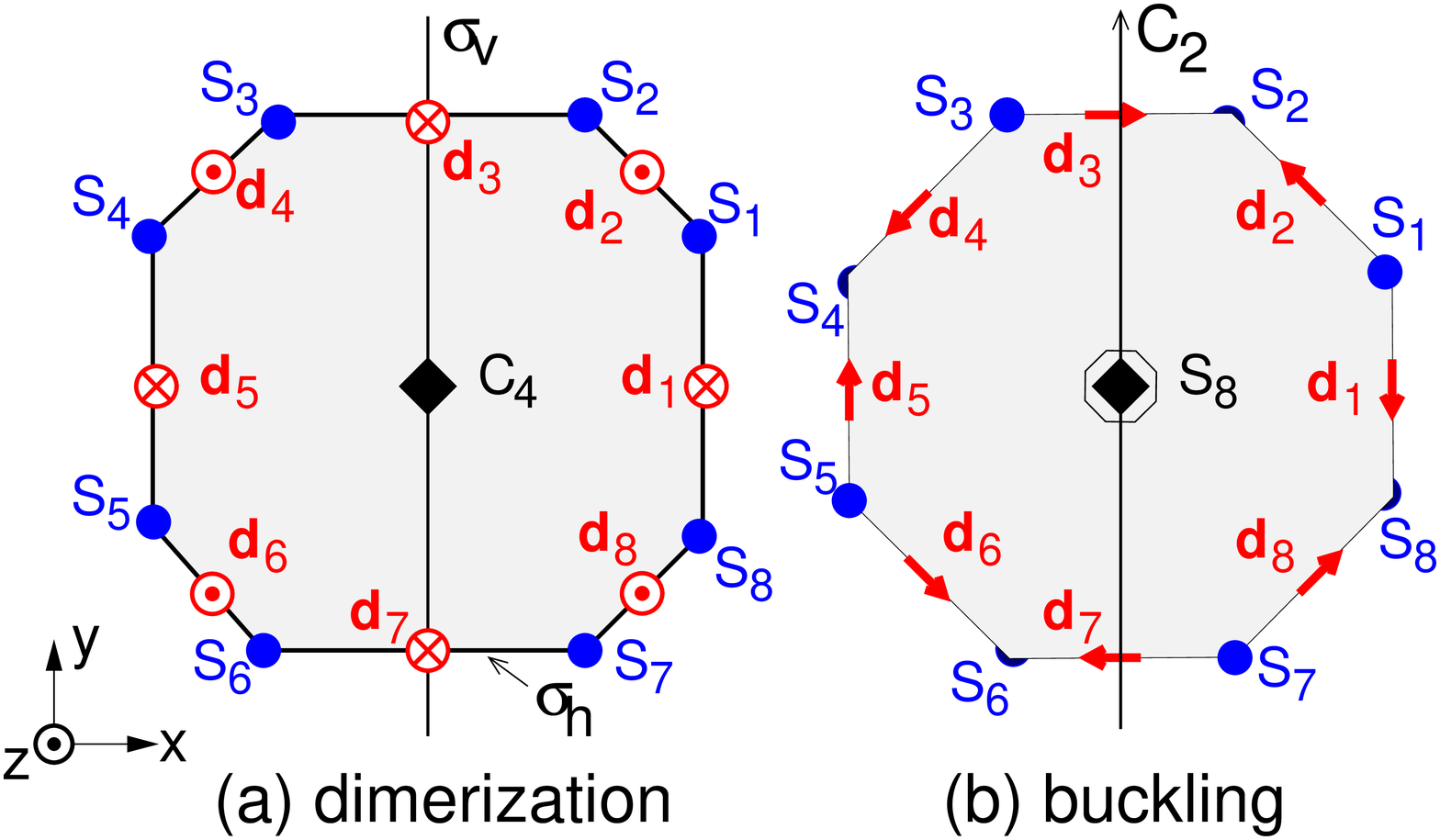}
\caption{\label{octagon.fig} 
(color on-line) Pictorial representation of the dimerized (a) and the buckled octagon (b).
The $\vec{d}$ vectors are determined by the symmetry generators. 
In (a) these are the $\mathsf{C}_4$ axis, the vertical reflection plane $\sigma_v$ and 
the horizontal plane $\sigma_h$ (here the generated group is $\mathsf{D}_{4h}=\mathsf{D}_4 \times \mathsf{i}$), while in (b) 
they are the rotoreflection $\mathsf{S}_8= \sigma_h \mathsf{C}_8$ and the $\mathsf{C}_{2}$ axis.}
\end{figure}

More recently, for the octanuclear CsFe8 cluster, a peculiar torque signal
was reported\cite{Waldmann1}, that appears quite abruptly around the lowest LC field.
This behaviour is clearly inconsistent with the previous mechanisms, and the
scenario of a field-induced magneto-elastic instability was suggested\cite{Waldmann1}.
In the same field regime, Schnelzer \textit{et al.}\cite{Schnelzer} found a large broadening of
the proton Nuclear Magnetic Resonance (NMR) spectrum which signals the
presence of large staggered transverse polarizations\cite{Schnelzer}, 
an effect which points to DM anisotropy\cite{Miyahara_06}.  
On the theoretical side, the experiments were first interpreted in the context of a
phenomenological 2-level approach\cite{Waldmann1,Waldmann2} which results, at the level of a single
molecule, in a structural instability around the LC due to an anisotropy 
induced off-diagonal coupling between the two lowest spin levels. 
This model successfully reproduced the experimental findings
qualitatively, but key questions such as which type of magnetic anisotropy
triggers the instability are beyond its reach. More recently, Soncini {\it et al.}\cite{Soncini}
discussed a model which also relies on phenomenological off-diagonal 
coupling parameters, but is based on elastic interactions between molecules 
and attributes the instability at the LC to the on-site magnetic anisotropy. 
The semi-empirical nature of these models calls for a microscopic approach 
which should also answer two fundamental questions: i) does a spontaneous structural
instability occur also in a microscopic model, and ii) what is the role of
(staggered) DM interactions, which should be expected to provide stronger
off-diagonal couplings than on-site anisotropy as they couple directly 
the two spin levels.

In this Letter, we investigate the possibility of a structural instability in the context of a fully microscopic model 
where a general $N$-site molecular wheel is described by a spin-$s$ Hamiltonian coupled to the lattice. 
This has allowed us to prove that molecular wheels are indeed unconditionally unstable at LCs.
For the case of CsFe$_8$, the general trends around the first LC  
show a remarkable agreement with all available experimental data\cite{Waldmann1,Schnelzer}   
for fields almost perpendicular to the plane of the ring, 
lending strong support to the present instability scenario. 
In addition, the model allows us to compare what happens at different LCs and field orientations 
and to make a number of predictions which go beyond the previous approaches. 
In particular, we find that the model is characterized by a much stronger tendency to distort at higher LCs.
It also shows a dramatic angular dependence whereby the torque shows a step-wise linear behavior 
or sharp anomalies for fields almost perpendicular or parallel to the ring plane respectively. 
The latter is found to be at least twice larger than what expected from the lowest order perturbation theory,
showing that sub-leading corrections play a substantial role.

The Hamiltonian we consider is adapted from the models used to describe the 
spin-Peierls transition in spin-1/2 chains, with two important differences:
(i) We add a single ion anisotropy since we are interested in spins larger than 1/2 (this term is also 
crucial to reproduce the characteristic background torque signal of Ref.~\cite{Waldmann1}), and 
(ii) we include a distortion-dependent DM interaction. The Hamiltonian reads:
\begin{eqnarray}\label{Ham.eq}
\mc{H}&=&\sum_{i} J_i \vec{s}_i \cdot \vec{s}_{i+1}  - D\sum_i s_{iz}^2 -\vec{B}\cdot\vec{S}\nonumber\\ 
&+&\frac{K}{2}\sum_i \delta_{i}^2 + \sum_{i} \vec{d}_{i} \cdot \vec{s}_i \times\vec{s}_{i+1}
\end{eqnarray} 
with implicit periodic boundary conditions. The nearest-neighbor exchange is antiferromagnetic
($J_i>0$) and the single-ion anisotropy easy-axis ($D > 0$).
$\delta_{i}$ denotes a deformation parameter of the bond (in the simplest case it is the variation of the length of the bond)
between the sites $i$ and $i+1$, and $\vec{d}_i$ is the corresponding DM vector, assumed to depend on $\delta_i$. 
In what follows, we define a fixed $xyz$ reference frame with the molecule on the $xy$-plane 
(cf. Fig.~\ref{octagon.fig}) and the field $\vec{B}$ in the $xz$-plane subtending an angle $\theta$ from the $z$-axis.

In the spin-Peierls transition of spin-1/2 chains, the structure factor diverges
at the zone boundary, and an instability can be triggered by an infinitesimal dimerization
of either the exchange integral\cite{CrossFisher} or the DM interaction if it is allowed by symmetry\cite{Zvyagin,Affleck}. Since the
exchange is typically much larger, its effect usually dominates. In the present case of a finite
system, the situation is quite different. In the absence of distortion, the spectrum is gapped 
except at the level crossings, and an infinitesimal distortion can lead to a structural instability
if and only if three conditions are met: (i) One sits at a level crossing, (ii) the induced perturbation
couples the two levels, and (iii) the coupling is of first order in the distortion to overcome the cost in elastic energy. 
Now, the invariance of the Hamiltonian under cyclic permutations $\mathsf{C}_{N}$ of the spin
indices implies that the eigenstates can be labeled by the momentum $k=2\pi n/N$ ($n=0,\ldots,N-1$) with 
the ground state alternating from $k=0$ to $k=\pi$ at subsequent LC fields\cite{perron}. 
By symmetry, the last two conditions can only be met by a dimerized DM interaction along $z$ 
(see Fig. \ref{octagon.fig}(a)). Indeed, a modulation of the exchange integrals 
(as studied e.g. in Ref.~\onlinecite{Spanu}) leads to an $\mathsf{SU}(2)$ invariant perturbation 
and cannot lift the degeneracy between states with different total spin, while the in-plane DM interactions
allowed by buckling (see Fig. \ref{octagon.fig}(b)) leads to a perturbation $\mc{V}_b$ which only couples the two levels to order $N/2$, 
and only when $N/2$ is even\cite{note1}. 
In addition, as for the exchange in the spin-Peierls mechanism, the dimerization of the DM interaction 
is expected to be linear in $\delta$ in the limit of small $\delta$ so that condition (iii) is met. 
This leads to the minimal set of conditions:
$J_{i}=J$, $\delta_{i}=(-1)^i\delta$, and $\vec{d}_i=(-1)^i d(\delta)~\vec{e}_z$, with $d(\delta)\propto \delta$
when $\delta \rightarrow 0$. All other effects are irrelevant in the limit of small $\delta_i$ and need not
be considered further given the purpose of the present paper. Our model is further specified by the following conventions. All energies are in units of $J=1$, 
while the scale of $\delta$ is implicitly set by the value of $K$.
Besides, we impose a cut-off on the DM interaction by choosing $d=d_0\tanh\delta$ with $d_0=0.05 J$.
Physically, such a cut-off must be imposed since DM interactions
are typically at most a few percent of the exchange. Its specific form has been chosen for numerical
convenience but does not affect the results qualitatively.

\begin{figure}[!t]
\includegraphics[width=0.4\textwidth]{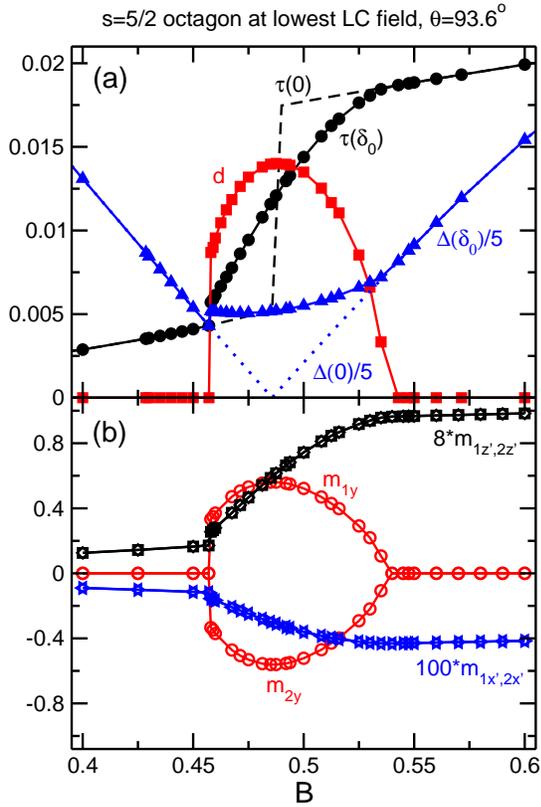}
\caption{\label{FirstLC_93p6.fig} 
(color on-line) Instability at the lowest LC field of CsFe$_8$  
for $\theta=93.6^\circ$ and $K=0.03$. 
(a) Optimal values of the lowest gap $\Delta$  
and the torque signal $\tau$ 
with and without the instability, and the DM amplitude $d=d_0 \tanh{\delta_0}$.  
(b) Local moments of any two neighboring sites (here 1 and 2, cf. Fig.~\ref{octagon.fig}(a)) in the rotated frame $x'yz'$ 
where $\vec{e}_{z'}$ is along $\vec{B}$. By symmetry, $m_{ix'}$ and $m_{iz'}$ are uniform while $m_{iy}$ is staggered.
As expected, $m_{iz'}\simeq 1/8$ after the LC field which corresponds to a total polarization of $S_{z'}\simeq 1$ shared by the 8 Fe$^{3+}$ ions.}
\end{figure}

The model is solved in two steps:
(i) Determine the ground state energy $E(\delta)$ for a given value of the distortion $\delta$. 
This has been achieved with exact diagonalizations based on a Lanczos algorithm; 
(ii) Find the distortion $\delta_0$ which minimizes the energy. This has been achieved by a bisection method. 
We have systematically investigated various $N$, $s$, $B$, $\theta$ and $K$. 
The main qualitative features are similar for all ring sizes and spins $s$. 
Here, in order to make contact with the reported experiments on CsFe$_8$, 
we restrict ourselves to the spin $s=5/2$ octagon case with $D=0.027$ 
(this corresponds to the experimental\cite{Waldmann1} estimates $D\simeq 0.56$ K and $J\simeq 20.6$ K for CsFe$_8$) 
and take $\theta=93.6^\circ, -3.3^\circ$ as considered in \cite{Waldmann1}.
As to the value of $K$, this is adjusted so that we get an approximate agreement with 
the width of the reported torque anomalies at large angles.   

First of all, let us note that the directions of the local magnetizations $\vec{m}_i=\langle \vec{s}_i\rangle$ 
and the torque $\boldsymbol{\tau}$ are fixed by symmetry. Indeed, 
the plane of $\vec{d}$ and $\vec{B}$ containing the center of the ring (cf. Fig.~\ref{octagon.fig}(a))
is a mirror plane\cite{note} which, together with the remaining $\mathsf{C}_4$ axis, 
gives $(m_{ix}, m_{iy}, m_{iz})=(m_{i+1,x}, -m_{i+1,y}, m_{i+1,z})$, i.e. 
the moments are uniform in the $xz$-plane and staggered along the $y$-axis. 
Accordingly, $\boldsymbol{\tau}$ points along the $y$-axis. 

Now, let us discuss the results we have obtained for the first
LC at $\theta=93.6^\circ$. 
As can be seen in Fig.~\ref{FirstLC_93p6.fig}(a), a finite distortion (and thus DM interaction) appears around the LC field. 
It opens a gap $\Delta$ and the jump in the torque $\tau$ is replaced by an almost linear behavior. 
The local magnetizations for (any) two neighboring spins are depicted in Fig. \ref{FirstLC_93p6.fig}(b). 
Apart from a linear-like uniform response in the $xz$-plane, we also find a large staggered magnetization 
along the $y$-axis with $|m_{iy}|\simeq 0.56$ at the center of the LC.
This should be contrasted with the phenomenological two-level model of \cite{Schnelzer} which 
gives a staggered response in the 
$xz$-plane (perpendicular to $\vec{B}$) which is $\sim 1.71$ at the center of the LC.
This difference stems from the phase of the off-diagonal coupling, which is purely imaginary in our
model and was assumed to be real in Ref.~\cite{Schnelzer}. 
A quantitative fit of the NMR data of \cite{Schnelzer} based on our results for 
the spin polarizations should take into account the anisotropic character of the dipolar hyperfine field,
which is clearly beyond the scope of the present work. 

We also note that the transition at the first critical field is weakly first order in our
model, which induces for example a small jump in the torque and in $m_{iy}$. 
This is not an artifact of the minimization method. Indeed, 
we have checked that the fourth-order derivative 
of the ground state energy functional $E(\delta_0)$ is negative in this parameter range.
This is not generic however, and the nature of the phase transitions 
as a function of $B$, $\theta$ and $K$ will be investigated in detail in a future article.

\begin{figure}[!tb]
\includegraphics[width=0.4\textwidth]{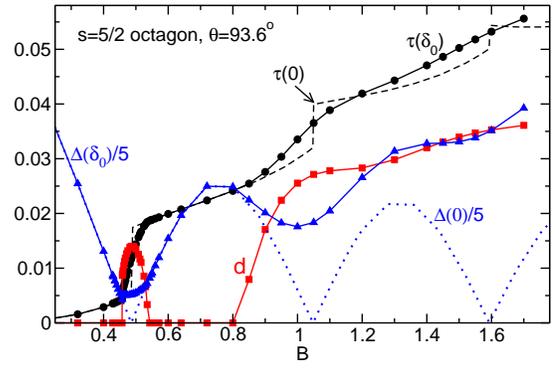}
\caption{\label{AllLCs_93p6.fig} 
(color on-line) Same as in Fig.\ref{FirstLC_93p6.fig}(a) but for a larger field interval which includes the lowest three LC points.}
\end{figure}

We now turn to the remaining LCs at $\theta=93.6^\circ$. 
Figure \ref{AllLCs_93p6.fig} shows the DM amplitude $d$ as well as the torque $\tau$ and the lowest energy gap $\Delta$, 
with and without the instability up to the third LC field. 
Quite remarkably, the tendency to distort becomes much larger
at higher LCs, actually so large that after the first critical field of the second
LC the distortion never disappears. This unexpected effect can be traced back to
the fact that the local spin polarizations in each $\vec{s}_i \times\vec{s}_j$ term  generally 
grow at large magnetizations (or fields), and by also noting that subsequent LC points are 
closer to each other at high fields due to the quasi-continuum character of the high energy spectrum\cite{note3}.
This tendency seems to be at variance with the assumption of Ref.~\cite{Soncini} that
the vibronic coupling decreases with the field. 
However, interestingly enough, this effect agrees with the general trend of the torque data at higher LCs,
where the almost linear behavior is smoothed out.

\begin{figure}[!t]
\includegraphics[width=0.4\textwidth]{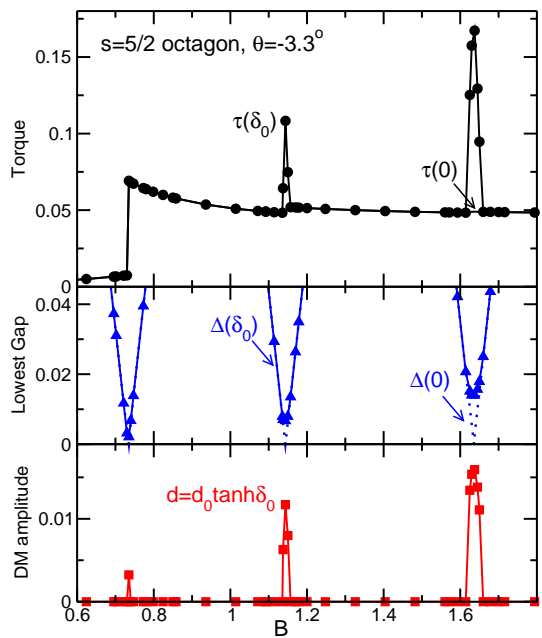}
\caption{\label{AllLCs_3p3.fig} 
(color on-line) Instability for the lowest three LCs of CsFe$_8$ for $\theta=93.6^\circ$ and $K=0.02$. 
(a) Torque $\tau$ and lowest gap $\Delta$ with and without the instability. (b) DM amplitude $d$.} 
\end{figure}

Finally, let us discuss the angular dependence of the torque (cf. Fig.~\ref{AllLCs_3p3.fig}). 
It turns out to be rather dramatic, in fact much more dramatic than anticipated by the phenomenological approach. 
For the set of parameters used at large angles, the instability for $\theta=-3.3^\circ$ at the first LC field could barely
be detected, and for the purpose of illustration, we have chosen a slightly smaller
stiffness $K$ to make the degeneracy lifting visible. 
This tendency is again in agreement with the experimental data of Ref.~\cite{Waldmann1}, where additional sharp torque anomalies
can be seen at the second and third LCs, but much weaker at the first one.
When comparing the shape of the anomalies with experiments, one should bear in mind that, for such narrow instabilities,
inhomogeneities are likely to lead to a distribution of critical fields broader than the instability
itself, leading to a smoothening and broadening of the anomalies.
A surprising aspect of our results is that while the shape of the distortion agrees with the dome-like 
contribution predicted by the lowest order perturbation theory,
their magnitude is at least twice as large\cite{note4}. This finding can be attributed to the sub-leading perturbative correction 
which scales as $d\cos\theta$\cite{Miyahara_06}. 
Finally, the dome-shape anomalies are expected to appear 
only when the step-height of the torque without the instability is sufficiently small. 
This explains why the sharp-anomalies are experimentally visible only at small angles $\theta$, except for the lowest LC field.

In conclusion, we have investigated the equivalent of the spin-Peierls instability in antiferromagnetic rings 
in the context of a microscopic spin model coupled to the lattice via a distortion dependent DM interaction.
Apart from demonstrating that magnetic rings are unstable at all LCs for an infinitesimal distortion, 
our results explain the general trends of all currently available data for the lowest LC of CsFe$_8$ at large and small angles, 
lending strong support to the present DM magnetoelastic model. 
Our model has also allowed us to make a number of specific predictions which go beyond the previous phenomenological approaches. 
For instance, we find that the instabilities are much stronger at high magnetic fields and may even persist from the second LC onward
for fields almost perpendicular to the ring plane.  
We also find sharply peaked torque anomalies when the field is almost parallel to the molecule in very homogeneous samples. 
It is our hope that the present paper will trigger further experimental investigation of these instabilities.

The authors are grateful to A. Parola, S. Bieri, M. Horvatic and C. Berthier 
for stimulating and enlightening discussions. 
This work was supported by the Swiss National Fund, by MaNEP, by the Hungarian 
OTKA Grants No.K62280 and No.K68807, and the ESF HFM project (Exchange Grant 1273).


\begin{thebibliography}{99}
\bibitem{GSV} D. Gatteschi, R. Sessoli, and J. Villain, \textit{Molecular Nanomagnets} (Oxford University Press, Oxford, 2006), and references therein.
\bibitem{Chudnovsky} E. M. Chudnovsky and J. Tejada, \textit{Macroscopic Quantum Tunneling of the Magnetic Moment},  
Cambridge Studies in Magnetism, Vol. 4 (Cambridge University Press, Cambridge, 1998), and references therein.
\bibitem{Chiorescu} I. Chiorescu, {\it et al.}, \prl {\bf 84}, 3454 (2000).
\bibitem{V6_PRL} I. Rousochatzakis, {\it et al.}, \prl{\bf 94}, 147204 (2005).
\bibitem{Affronte} M. Affronte, {\it et al}, \prl{\bf 88}, 167201 (2002). 
\bibitem{DM} I. E. Dzyaloshinskii, J. Phys. Chem. Solids, {\bf 4}, 241 (1958); T. Moriya,  \prl {\bf 4}, 228 (1960).
\bibitem{Cinti} F. Cinti, {\it et al.}, Eur. Phys. J. B {\bf 30}, 461-468 (2002).
\bibitem{Miyashita} H. Nakano and S. Miyashita, J. Phys. Soc. Jpn. {\bf 71}, 2580 (2002).
\bibitem {Waldmann1} O. Waldmann, {\it et al.}, \prl {\bf 96}, 027206 (2006).
\bibitem{Schnelzer} L. Schnelzer, {\it et al.},  \prl{\bf 99}, 087201 (2007).
\bibitem{Miyahara_06} S. Miyahara, {\it et al.}, \prb{\bf 75}, 184402 (2007).
\bibitem {Waldmann2} O. Waldmann, \prb {\bf 75}, 174440 (2007).
\bibitem {Soncini} A. Soncini and L. F. Chibotaru, \prl {\bf 99}, 077204 (2007).
\bibitem{CrossFisher} M. C. Cross and D. S. Fisher, Phys. Rev. B {\bf 19}, 402 (1979).        
\bibitem{Zvyagin} A. A. Zvyagin, Zh. Eksp. Teor. Fiz. {\bf 98}, 1396-1401 (1990).
\bibitem{Affleck} I. Affleck and M. Oshikawa, Phys. Rev. B {\bf 60}, 1038 (1999).
\bibitem{perron}  W. Marshall, Proc. R. Soc. (London) {\bf A232 }, 48 (1955);  E. Lieb and D.C. Mattis, J. Math. Phys. {\bf 3}, 749 (1962).
\bibitem{Spanu} L. Spanu and A. Parola, \prl{\bf 92}, 197202 (2004).
\bibitem{note1} Since $\mathsf{C}_{N}^{N/2} \mc{V}_b = \mp\mc{V}_b$ for $N/2$ even or odd,
$\mc{V}_b$ contains representations with $n=1,3,\ldots,N-1$ for $N/2$ even, and $n=0,2,\ldots,N-2$ for $N/2$ odd. 
Thus a $k=\pi$ momentum transfer necessary for the level mixing will first appear in $N/2$-th order in $d$ and only for $N/2$ even.  
\bibitem{note} Here the mirror operation stands for a reflection through the $xz$-plane followed by the time reversal operation $\mc{T}$.
\bibitem{note3} For CsFe$_8$, we have checked numerically that the off-diagonal coupling $v_j$ between the lowest levels 
at the $j$-th LC field grow with $j$. For instance, at  $\theta=-3.3^\circ$, $v/d\simeq(0.1565, 0.2858, 0.4291)$ 
at the lowest three LC fields $B_c\simeq(0.734, 1.145, 1.6373)$, while at $\theta=93.6^\circ$, $v/d\simeq (0.9251, 1.6327, 2.3154)$ 
at the corresponding fields $B_c\simeq (0.485, 1.052, 1.598)$. 
For a spin-$s$ tetramer, $v_j=j \sqrt{ \frac{j}{8 (2 j+1)} \left( (4 s+1)^2-j^2\right) } d \sin \theta $.
\bibitem{note4} For instance, for the second and the third LC field at $\theta=-3.3^\circ$,  
lowest order perturbation theory predicts $d_{\text{max}} \simeq (0.0045, 0.0067)$ 
to be compared with the corresponding values $(0.0117, 0.0159)$ (cf. Fig.~\ref{AllLCs_3p3.fig}). 
\end{thebibliography}
\end{document}